\documentclass[pra,a4paper,twocolumn,superscriptaddress,longbibliography,nofootinbib]{revtex4-2}
\usepackage[utf8]{inputenc}
\usepackage{amssymb}
\usepackage{amsmath}
\usepackage{amsfonts}
\usepackage{amsthm}
\usepackage{amsbsy}
\usepackage{bm}
\usepackage{bbold}

\usepackage{graphicx}
\usepackage{xcolor}
\usepackage{multirow}
\usepackage{float}
\usepackage{comment}
\usepackage{ulem}
\usepackage{relsize}
\usepackage{cmap}
\usepackage{fancyhdr}
\usepackage[backref = true,colorlinks=true]{hyperref}

\begin{document}

\title{Diffusive modes of two-band fermions under number-conserving dissipative dynamics}

\author{A.\,A.\, Lyublinskaya}

\affiliation{\mbox{L. D. Landau Institute for Theoretical Physics, Semenova 1-a, 142432, Chernogolovka, Russia}}

\affiliation{Moscow Institute for Physics and Technology, 141700, Moscow, Russia}

\author{I.\,S.\, Burmistrov}

\affiliation{\mbox{L. D. Landau Institute for Theoretical Physics, Semenova 1-a, 142432, Chernogolovka, Russia}}

\affiliation{Laboratory for Condensed Matter Physics, HSE University, 101000, Moscow, Russia}


\begin{abstract}
Driven-dissipative 
protocols are proposed to control and create nontrivial quantum many-body correlated states. Protocols conserving the number of particles stand apart. As well-known, in quantum systems with the unitary dynamics the particle number conservation and random scattering yield diffusive behavior of two-particle excitations (diffusons and cooperons). Existence of diffusive modes in the particle-number-conserving dissipative dynamics is not well studied yet. We explicitly demonstrate the existence of diffusons in a paradigmatic model of a two-band system, with dissipative dynamics aiming to empty one fermion band and to populate the other one. The studied model is generalization of the model introduced in F. Tonielli, J. C. Budich, A. Altland, and S. Diehl, Phys. Rev. Lett. 124, 240404 (2020). We find how the diffusion coefficient depends on details of a model and the rate of dissipation. We discuss how the existence of diffusive modes complicates engineering of macroscopic many-body correlated states. 
\end{abstract}


\maketitle

\thispagestyle{fancy}
\fancyhead{}
\fancyhead[LO,CE]{\textsf{\footnotesize submitted to JETP Letters}}
\fancyhead[RO,RE]{1}
\fancyfoot{}

\noindent{D}issipative dynamics of open quantum many-body systems has recently attracted a lot of interest~\cite{Diehl2016,LeHur2018,Skinner2019,Rudner2020,Thompson2023} because it paves the way to deal with non-equilibrium states of matter, theoretically ~\cite{Lechner2013,Piazza2014,Keeling2014,Altman2015,Kollath2016} and experimentally~\cite{Devoret2015}. This area of research abounds with non-trivial and non-intuitive results, including, for instance, non-equilibrium phase transitions~\cite{DallaTorre2010,Sieberer2013,Raftery2014,Li2018,Li2019,Roy2020,Garratt2021}.  
Driven-dissipative preparations allow one to create in a controlled way quantum many-body correlated steady states that have no analogue in the Hamiltonian dynamics~\cite{Diehl2008,Kraus2008,Verstraete2009,Weimer2010,Diehl2011,Bardyn2012,Bardyn2013,Otterbach2014,Konig2014,Lang2015,Budich2015,Iemini2016,Zhou2017,Gong2017,Goldstein2019,Shavit2020,Tonielli2020,Yoshida2020,Gau2020a,Gau2020b,Bandyopadhyay2020,Santos2020,Altland2021,Beck2021,Nava2023,Shkolnik2023}.

One of the most common analytical methods for describing quantum systems subjected to external source of dissipation is Gorini-Kossakovski-Sudarshan-Lindblad (GKSL) master equation \cite{Lindblad1976,GKS1976} in which the dynamics of the density matrix is explicitly divided into unitary and dissipative parts, defined by the Hamiltonian and jump operators, respectively. Recently, the mapping of GKSL equation to the quantum field theory on Keldysh time contour has been developed (see Refs. \cite{Diehl2016,Thompson2023} for a review). 

As well-known, symmetries and conservation laws are guiding principles for the quantum field theory. For Hamiltonian systems the particle number conservation in combination with random scattering results in diffusive dynamics. Also diffusion can appear due to dephasing caused by a coupling to a bath \cite{Esposito2005,Castro-Alvaredo2016,Dhar2019,Jin2022}. 
How does diffusion emerge in driven-dissipative preparations conserving the number of particles? It is not established so far. One of obstacles is that dissipative state preparation is constructed in such a way that avoids randomness in GKSL master equation. 

In this Letter, we address 
a general question of 
existence of diffusion modes in 
particle-number-conserving dissipative dynamics.
Do diffusive two-particle excitations (diffusons and cooperons), which are familiar for disordered Hamiltonians, exist in number-conserving dissipative systems described by GSKL master equation?
We show that in a wide class of number-conserving dissipative systems
diffusion occurs naturally, as it 
occurs under the unitary evolution.

\begin{figure}[t]
\centerline{\includegraphics[width=0.9\columnwidth]{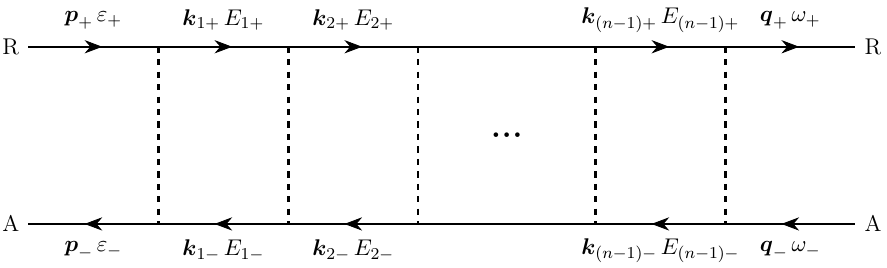}
}
\caption{Ladder diagram for the diffuson. The solid lines denote the self-consistent Green's function. The dashed lines indicate the dissipation-induced interaction (the boson Green's function $\mathcal{D}_{\bm{p},\textsf{a},\alpha}(\omega)$). 
}
\label{Fig:Ladder}
\end{figure}

To be specific, we consider a generalization of the model studied in Refs. \cite{Tonielli2020,Nosov2023}. 
It is a two-band fermions, which are scattered off by random dynamical bosonic fields serving as a quantum noise. {\color{black} In contrast to Refs. \cite{Tonielli2020,Nosov2023}, our generalized model allows for non-local scattering.}
Averaging over quantum noise manifests itself as dissipative dynamics within GKSL equation with jump operators transferring the fermion population from the upper band to the lower one (and vice versa)
with the rate determined by scattering on bosonic fields.
Calculation of the sum of 
ladder-type diagrams (diffuson) 
with the dissipation-induced interaction lines, see Fig. \ref{Fig:Ladder},  leads to the canonical-type expression with a diffusion pole, cf. Eq. \eqref{eq:Diff:f}. The corresponding diffusion coefficient depends on parameters of the model, cf. Eq. \eqref{eq:res:D}.


\noindent\textsf{\color{blue} Model. --- } We consider the following partition function on the Keldysh contour:
\begin{equation}
Z[\overline{\Phi},\Phi]= \int \mathcal{D}[\overline{\psi}_\pm,\psi_\pm]\,  e^{i S_{\rm 0} + i S_{\rm \Phi}} .
\label{eq:K:action}
\end{equation}
It depends on auxiliary bosonic fields $\Phi$ and $\overline{\Phi}$. 
Here $\psi_{\pm}{=}\{\psi_{1,\pm},\psi_{2,\pm}\}$ ($\overline{\psi}_{\pm} {=}\{\overline{\psi}_{1,\pm},\overline{\psi}_{2,\pm}\}$) denote spin $s{=}1/2$ fermionic fields, corresponding to annihilation and creation operators, on the forward (`+') and backward (`-') contour. The kinetic part of the action describes a free electron gas,\begin{gather}\label{eq:Action:Hamiltonian}
S_{\rm 0}  {=} \int\limits_{\bm{q},t}\sum_{\tau=\pm} \tau \overline{\psi}_{\bm{q},\tau}(t) \bigl(  i \partial_t {-} H_{\rm 0}(\bm{q}) \bigr ) \psi_{\bm{q},\tau}(t) ,
\end{gather}
where $\bm{q}$ is the $d$-dimensional momentum, and we use a shorthand notation: $\int_{\bm{q},t}{\equiv} \int dt \int d^d\bm{q}/(2\pi)^d$. The $2{\times}2$ Hamiltonian $H_{\rm 0}(\bm{q})$ acts in the spin space. 
We assume that $H_{\rm 0}(\bm{q})$ can be diagonalized by a unitary $2{\times}2$ matrix $U_{\bm{q}}$ such that $H_{\rm 0}(\bm{q}){=} \xi_q U_{\bm{q}} \sigma_z U^\dag_{\bm{q}}$. Here $\sigma_z{=}{\rm diag}\{1,{-}1\}$ stands for the Pauli matrix.
 It is convenient to introduce another set of fermionic fields $c_{\bm{q}} {=} \{c_{\bm{q},\textsf{u}},c_{\bm{q},\textsf{d}}\}{=} U^\dag_{\bm{q}} \psi_{\bm{q}}$ and $\overline{c}_{\bm{q}} {=} \{\overline{c}_{\bm{q},\textsf{u}},\overline{c}_{\bm{q},\textsf{d}}\}{=} \overline{\psi}_{\bm{q}} U_{\bm{q}}$. These fermionic fields correspond to creation and annihilation of fermions in the `up' (with energy ${+}\xi_q$) and `down' (with energy ${-}\xi_q$) bands, such that
\begin{gather}\label{eq:Action:Hamiltonian:C}
S_{\rm 0}  {=} \int\limits_{\bm{q},t}\sum_{\tau=\pm} \tau \overline{c}_{\bm{q},\tau}(t) \bigl(  i \partial_t {-} \xi_{\bm{q}}\sigma_z \bigr ) c_{\bm{q},\tau}(t) .
\end{gather}

The free fermions experience scattering off random dynamical fields $\Phi$ and $\overline{\Phi}$ which are described by the following  action local in space and time,
\begin{align}
S_{\rm \Phi} {=} & \int\limits_{\bm{x},t} \sum_{\alpha{=}1,2}\sum_{\textsf{a}{=}\textsf{u},\textsf{d}}\sum_{\tau{=}\pm}\tau \Bigl\{
\overline{l}_{\textsf{a},\tau}(\bm{x},t{-}\delta^\prime \tau)\overline{\Phi}_{\textsf{a},\alpha,\tau}(\bm{x},t)\notag \\
& {\times} \psi_{\alpha,\tau}(\bm{x},t{-}\delta^\prime \tau{-}\epsilon^\prime_\textsf{a}\tau) 
 {+} \overline{\psi}_{\alpha,\tau}(\bm{x},t{-}\delta^\prime\tau) \Phi_{\textsf{a},\alpha,\tau}(\bm{x},t)\notag \\
 & {\times}  
l_{\textsf{a},\tau}(\bm{x},t{-}\delta^\prime\tau{-}\epsilon^\prime_\textsf{a}\tau) ,
\Bigr \} ,
\label{eq:def:S:Phi}
\end{align}
where $\int_{\bm{x},t}{\equiv} \int dt \int d\bm{x}$. Here we introduce the other set of fermionic fields $\overline{l}_{\bm{q},\textsf{a}}{=}v^*_{\bm{q}}\overline{c}_{\bm{q},\textsf{a}}$ and $l_{\bm{q},\textsf{a}}{=} v_{\bm{q}} c_{\bm{q},\textsf{a}}$, where  $v_{\bm{q}}$ is an auxiliary function of momentum. Scattering of fermions on bosonic fields described by Eq. \eqref{eq:def:S:Phi} is unusual, since during the scattering, fermions transform from one basis to the other. In virtue of the relation $l_{\bm{q}}{=} v_{\bm{q}} U_{\bm{q}}^\dag \psi_{\bm{q}}$, 
the scattering of bosonic fields $\psi$ becomes effectively non-local in space. However, for a given matrix $U_{\bm{q}}$, there is a particular choice of $v_{\bm{q}}$ that makes the scattering to be spatially local. 
{\color{black} 
Although Refs. \cite{Tonielli2020,Nosov2023} were focused on such a situation, as we shall demonstrate below, this is not necessary for appearance of diffusion.}
Also we emphasize that the scattering in Eq. \eqref{eq:def:S:Phi} conserves the total number of particles. 

The random dynamical bosonic fields $\Phi$ and $\overline{\Phi}$ in Eq. \eqref{eq:def:S:Phi} are assumed to be Gaussian, uncorrelated in space and time, and with zero mean. The only non-zero pair correlation functions are as follows, 
\begin{align}
\langle \Phi_{\textsf{a},\alpha,\pm}(\bm{x},t)\overline{\Phi}_{\textsf{a},\alpha,\pm}(\bm{x^\prime},t^\prime)\rangle & {=}\gamma_\alpha^{(\textsf{a})} 
\delta(\bm{x}{-}\bm{x^\prime})\delta(t{-}t^\prime{\pm}\delta_\textsf{a}) ,\notag \\
\langle \Phi_{\textsf{a},\alpha,\tau_\textsf{a}}(\bm{x},t)\overline{\Phi}_{\textsf{a},\alpha,{-}\tau_\textsf{a}}(\bm{x^\prime},t^\prime)\rangle & {=}\gamma_\alpha^{(\textsf{a})} 
\delta(\bm{x}{-}\bm{x^\prime})[\delta(t{-}t^\prime{+}\delta_\textsf{a})
\notag \\
& {+}\delta(t{-}t^\prime{-}\delta_\textsf{a})] ,
\label{eq:def:PhiPhi}
\end{align}  
where $\tau_{\textsf{u/d}}{=}{\pm}1$ and $\gamma_\alpha^{(\textsf{a})}$ is the rate of scattering between a fermion in state with the spin projection $\alpha{=}1,2$ and a fermion in the band  $\textsf{a}{=}\textsf{u},\textsf{d}$. In Eqs. \eqref{eq:def:S:Phi} and \eqref{eq:def:PhiPhi} we introduced a number of equal-time regulators: $\epsilon^\prime_{\textsf{u}}{=}{-}\epsilon^\prime_{\textsf{d}}{=}{-}\epsilon^{\prime}{>}0$, $\delta_{\textsf{u}}{=}{-}\delta_{\textsf{d}}{=}\delta^{\prime\prime}{>}0$.  They satisfy the inequality $\delta^\prime{>}\delta^{\prime\prime}{>}\epsilon^\prime{>}0$. At the end of the calculations, we set them all to zero. 

It is worthwhile to mention that after the Keldysh rotation \cite{Kamenev2009}, the bosonic pair correlation function acquires a standard structure in the Keldysh space:
\begin{gather}
\mathcal{D}_{\bm{k},\textsf{a},\alpha}(\omega) =
-i\langle \Phi_{\textsf{a},\alpha,{\rm cl/q}}(\bm{x},t)\overline{\Phi}_{\textsf{a},\alpha,{\rm cl/q}}(\bm{x^\prime},t^\prime)\rangle_{\bm{k},\omega}\notag \\
=
 i \gamma_\alpha^{(\textsf{a})} s_\textsf{a}\begin{pmatrix}
-2 s_\textsf{a} &  1_\omega^R \\
 - 1_\omega^A & 0
\end{pmatrix} ,
\label{eq:def:PhiPhi:RAK}
\end{gather}
where $s_\textsf{u}{=}{-}s_\textsf{d}{=}1$. The superscript $R/A$ indicates that $1_\omega^{R/A}$ corresponds to $\delta(t\mp 0^+)$ after the Fourier transform to time domain. Also, Eq. \eqref{eq:def:PhiPhi:RAK} indicates that the random bosonic fields have the distribution function equal $-s_\textsf{a}$. Therefore, the bosonic fields (in a frequency range of interest) correspond to the equilibrium bath at $T{=}0$.

\noindent\textsf{\color{blue} Keldysh action in terms of $c$-fermions. --- } The partition function \eqref{eq:K:action} is a random quantity with some distribution function, whose computation is an interesting problem. In this work, we restrict ourselves to study the average partition function only, 
\begin{gather}
\langle Z[\Phi,\overline{\Phi}]\rangle_{\Phi} = \int \mathcal{D}[\overline{\Psi},\Psi]\,  e^{i S_{\rm 0} + i S_{\rm L}}, \quad  S_{\rm L} = \frac{i}{2}\langle S_{\rm \Phi}^2\rangle_{\Phi}
. 
\end{gather}
 The Keldysh action $S_L$ can be explicitly written in the basis of $c$-fermions as
\begin{align}
S_{\rm L} &{=} {-} i (2\pi)^d \!\!\int\limits_{\bm{p_{j}},t}\! \delta(\bm{p_1}{-}\bm{p_2}{+}\bm{p_3}{-}\bm{p_4}) 
\sum_{\textsf{a}=\textsf{u},\textsf{d}}\sum_{\alpha=1,2}  \gamma_\alpha^{(\textsf{a})}\sum_{\tau{=}\pm}
\notag \\
{\times} &
\Bigl [ \overline{c}_{\bm{p_1},-}(t) \overline{\mathcal{L}}^{(\textsf{a},\alpha)}_{\bm{p_1p_2}}
c_{\bm{p_2},-}(t^{+}_\textsf{a})
\overline{c}_{\bm{p_3},+}(t{-}\delta)\mathcal{L}^{(\textsf{a},\alpha)}_{\bm{p_3p_4}}
c_{\bm{p_4},+}(t^{-}_\textsf{a}{-}\delta)
\notag \\
{-} &
 \overline{c}_{\bm{p_1},\tau}(t^{\tau}_\textsf{a})\overline{\mathcal{L}}^{(\textsf{a},\alpha)}_{\bm{p_1p_2}}
c_{\bm{p_2},\tau}(t)
\overline{c}_{\bm{p_3},\tau}(t{-}\delta_\tau)\mathcal{L}^{(\textsf{a},\alpha)}_{\bm{p_3p_4}}
c_{\bm{p_4},\tau}(t^{\tau}_\textsf{a}{-}\delta_\tau)
\Bigr ] .
\label{eq:SL:ud}
\end{align}
Here we also used equal-time regularization $t^{\pm}_\textsf{a}{=}t{\pm}\epsilon_\textsf{a}$ and $\delta_\tau{=}\tau \delta$, where $\epsilon_{\textsf{u}}{=}{-}\epsilon_{\textsf{d}}{=}{-}\epsilon$ and $\delta{>}\epsilon{>}0$. The four matrices $\mathcal{L}^{(\textsf{a},\alpha)}$ act in the $\textsf{u}/\textsf{d}$ space and are defined as follows
\begin{gather}
[\mathcal{L}^{(\textsf{u},\alpha)}_{\bm{pq}}]_{\textsf{ab}} =v_{\bm{q}} [U_{\bm{p}}^\dag]_{\textsf{a},\alpha}\delta_{\textsf{bu}}, \quad   [\mathcal{L}^{(\textsf{d},\alpha)}_{\bm{pq}}]_{\textsf{ab}}  = -v_{\bm{p}}^* [U_{\bm{q}}]_{\alpha,\textsf{b}}\delta_{\textsf{ad}} ,\notag \\
\overline{\mathcal{L}}_{\bm{pq}}^{(\textsf{a},\alpha)}=[\mathcal{L}_{\bm{qp}}^{(\textsf{a},\alpha)}]^\dag .
\label{eq:A:matrices:0}
\end{gather}
The action $S_{\rm 0}{+}S_{\rm L}$ is invariant under global $\mathrm{U}(1){\times} \mathrm{U}(1)$ transformations, $\overline{c}_{\pm}{\to} e^{-i\chi_\pm} \overline{c}_{\pm}$ and $c_{\pm}{\to} e^{ i\chi_\pm} c_{\pm}$ with $\chi_+{\neq}\chi_{-}$. This strong symmetry is a manifestation of the conservation of the total number of particles in the model. We emphasize that there is neither strong nor weak $U(1)$ symmetry of the action that could be associated with the conservation of $\textsf{u}$- or $\textsf{d}$-fermions separately. 
The translation invariance is a weak symmetry of the model \cite{Diehl2016,Buca2012,Albert2014}, since $S_{\rm L}$ is invariant under translations $\overline{c}_{\bm{q},\pm}{\to} e^{-i\bm{q}\bm{\eta}_\pm} \overline{c}_{\bm{q},\pm}$ and $c_{\bm{q}\pm}{\to} e^{i\bm{q}\bm{\eta}_\pm} c_{\bm{q},\pm}$ with $\bm{\eta}_+{=}\bm{\eta}_-$ only, i.e. acting identically on the forward and backward branches of the Keldysh contour. We note that if bosonic fields were a real random potential, then even weak symmetry for translation invariance would be absent.
However, absence of strong symmetry for translation invariance makes possible for $c$-fermions to transfer not only energy but also to relax momentum during scattering on the bosonic fields.

\noindent\textsf{\color{blue} Master equation. --- } 
We note that the average partition function $\langle Z[\Phi,\overline{\Phi}]\rangle_{\Phi} $ corresponds to the density matrix $\rho$ governed by the following GKSL master equation,
\begin{gather}
\frac{d\rho}{dt} {=} \int\limits_{\bm{x}} 
\Bigl ( i [\rho,H_{\rm 0}] {+}\sum_{\textsf{a}=\textsf{u},\textsf{d}}\sum_{\alpha=1,2} \gamma_\alpha^{(\textsf{a})}
\bigl (2 L_{\textsf{a},\alpha} \rho L_{\textsf{a},\alpha}^\dag \notag \\
{-} \{L_{\textsf{a},\alpha}^\dag L_{\textsf{a},\alpha}, \rho\}\bigr )\Bigr ) ,
\label{eq:GKSL}
\end{gather}
where the jump operators are given as $L_{\textsf{u},\alpha} {=} \psi^{\dag}_{\alpha}(\bm{x})  l_{\textsf{u}}(\bm{x})$ and $L_{\textsf{d},\alpha} {=} \psi_{\alpha}(\bm{x}) l^{\dag}_{\textsf{d}}(\bm{x})$. We note that the scattering rates $\gamma_\alpha^{(\textsf{a})}$, in fact, determine the rates of dissipation. 
In the case of half-filling, the GKSL equation has a steady state solution --- the dark state, $\rho{=}|D\rangle \langle D|$, in which the $\textsf{d}$-band is fully occupied while the $\textsf{u}$-band is empty. 

\begin{figure}[t]
\centerline{\includegraphics[width=0.8\columnwidth]{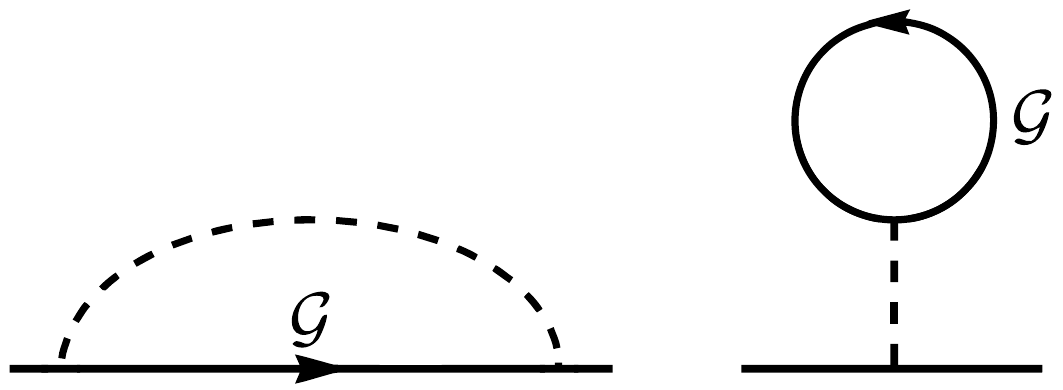}
}
\caption{Self-energy diagrams of the Fock- and Hartree-type in the self-consistent Born approximation. The solid lines denote the self-consistent Green's function. The dashed curve indicates the dissipation-induced interaction (the boson Green's function $\mathcal{D}_{\bm{p},\textsf{a},\alpha}(\omega)$). 
}
\label{Fig:SCBA}
\end{figure}

\noindent\textsf{\color{blue} Self-consistent Born approximation. --- } The dark state can be obtained from the analysis of the Keldysh action $S_{\rm 0}{+}S_{\rm L}$. It corresponds to the self-consistent solution of the Dyson equation written for the single-particle Green's function in the lowest order in the dissipation strength $\gamma_\alpha^{(a)}$ (see Fig. \ref{Fig:SCBA}) \cite{Tonielli2020}. 

The self-consistent Green's functions are diagonal in the $\textsf{u/d}$-space and given as (see Ref. \cite{Nosov2023} for details)
\begin{equation}
\begin{split}
\mathcal{G}^{R/A}_{\bm{q},\textsf{a}}(\varepsilon) & = \Bigl [\varepsilon-\xi_q s_\textsf{a} \pm i \bar{\gamma}_\textsf{a} |v_q|^2\Bigr ]^{-1} , \\ \mathcal{G}^K_{\bm{q},\textsf{a}}(\varepsilon) & = 
 s_\textsf{a} \Bigr [ \mathcal{G}^{R}_{\bm{q},\textsf{a}}(\varepsilon)-\mathcal{G}^{A}_{\bm{q},\textsf{a}}(\varepsilon) \Bigl ] ,
 \end{split}
 \label{eq:G:SCBA}
\end{equation}
where $\bar{\gamma}_\textsf{a}{=} \int_p [U^\dag_{\bm{p}} \hat\gamma^{(\textsf{a})} U_{\bm{p}}]_{\textsf{a}\textsf{a}}$.
Here we introduce $2{\times}2$ matrix $\hat\gamma^{(\textsf{a})} {=} {\rm diag}\{\gamma^{(\textsf{a})}_1,\gamma^{(\textsf{a})}_2\}$. We note that the factor $s_\textsf{a}{=}{\pm} 1$ determines the distribution function (as $(1{-}s_\textsf{a})/2$) of the $c$-fermions in the up and down bands. Therefore, Eq. \eqref{eq:G:SCBA} describes indeed the dark state with the fully occupied $\textsf{d}$-band and the completely empty $\textsf{u}$-band.

\noindent\textsf{\color{blue} Ladder summation for diffuson. --- } As well-known, diffusion of particles in disordered systems corresponds to diffuson which is a particle-hole excitation described by the impurity scattering ladder diagrams \cite{Lee1985}. Since  there is momentum relaxation in the model considered, it is natural to expect that ladder diagrams with dissipative lines (the bosonic correlation function \eqref{eq:def:PhiPhi}) could produce diffusion pole. Let us consider the 
two-particle irreducible average 
\begin{equation}
\langle\!\langle c_{\bm{p}_+,\textsf{a},\nu}(t_1) \bar{c}_{\bm{p}_-,\textsf{b},\mu}(t_1)\cdot c_{\bm{q}_-,\textsf{b}^\prime,\mu^\prime}(t_2)  \bar{c}_{\bm{q}_+,\textsf{a}^\prime,\nu^\prime}(t_2)\rangle\!\rangle  ,
\label{eq:2PC}
\end{equation}
 where $\bm{p}_\pm {=} \bm{p}{\pm} \bm{Q}/2$, $\bm{q}{=}\bm{q}{\pm}\bm{Q}/2$,
and $\nu,\nu^\prime,\mu,\mu^\prime{=}1,2$ are indices in the rotated Keldysh space. Such a two-particle irreducible average corresponds to the density-density correlation function and in the absence of dissipation produces the product of two Green's functions, $\mathcal{G}^{\nu\nu^\prime}_{\bm{p}_+,\textsf{a}}(\varepsilon_+) 
 \mathcal{G}^{\mu^\prime \mu}_{\bm{p}_-,\textsf{b}}(\varepsilon_-)\delta_{\textsf{a}\textsf{a}^\prime} \delta_{\textsf{b}\textsf{b}^\prime}\delta(\bm{p}{-}\bm{q})$,  (after Fourier transform from the time domain to the frequency one) at the level of self-consistent Born approximation, where $\varepsilon_\pm{=}\varepsilon{\pm} \Omega/2$. We note that we do not consider the ladder for cooperon which is generated by correlation function \eqref{eq:2PC} 
 with $\bar{c}(t_1)$ substituted by $c(t_1)$ and ${c}(t_2)$ changed to $\bar{c}(t_2)$. Due to the equal-time dissipation-induced interaction in $S_{\rm L}$, the cooperon ladder vanishes.
 
In order to treat the two-particle correlation function \eqref{eq:2PC} beyond the self-consistent Born approximation, it is convenient to rewrite the action $S_{\rm L}$ in the rotated Keldysh basis. We note that for computation of the ladder, it is not needed to keep track of the equal-time regularization. Then we obtain
\begin{gather}
S_{\rm L} {=} \frac{i}{2} \gamma (2\pi)^d  \!\!  \int\limits_{\bm{p_{j}},t}\!
\delta(\bm{p_1}{-}\bm{p_2}{+}\bm{p_3}{-}\bm{p_4})
\sum_{\textsf{a}{=}\textsf{u},\textsf{d}} \sum_{\alpha{=}1,2} \sum_{\nu,\mu{=}0,1}
\notag \\
\times P_{\mu\nu}
\overline{c}_{\bm{p_1}}(t) \tau_\mu \overline{\mathcal{L}}^{(\textsf{a},\alpha)}_{\bm{p_1p_2}} c_{\bm{p_2}}(t)
\
\overline{c}_{\bm{p_3}}(t) \tau_\nu\mathcal{L}^{(\textsf{a},\alpha)}_{\bm{p_3p_4}}c_{\bm{p_4}}(t) .
 \end{gather} 
 Here $\tau_{0}$ and $\tau_1$ are the identity matrix and the standard $\tau_x$ Pauli matrix, respectively. They act in the Keldysh space. Also, we introduced $2{\times}2$ matrix $P$ with the following matrix elements, 
$P_{00}{=}2$, $P_{01}{=}{-}P_{10}{=}1$, and $P_{11}{=}0$.

In addition,  for computation of the ladder diagrams it is convenient to write the self-consistent Green's function as $\mathcal{G}_{\bm{p},\textsf{a}}(\varepsilon){=}\mathcal{G}^R_{\bm{p},\textsf{a}}(\varepsilon) \Lambda^{(+)}_\textsf{a} {+} \mathcal{G}^A_{\bm{p},\textsf{a}}(\varepsilon) \Lambda^{(-)}_\textsf{a}$ where
 \begin{equation}
\Lambda^{({+})}_\textsf{a}=\begin{pmatrix}
1 & s_\textsf{a} \\
0 & 0 
\end{pmatrix} , \quad 
\Lambda^{(-)}_\textsf{a}=\begin{pmatrix}
0 & -s_\textsf{a} \\
0 & 1 
\end{pmatrix} .
\end{equation} 
We note that matrices $\Lambda^{({\pm})}$ are orthogonal projectors, $\Lambda^{({+})}\Lambda^{({-})}{=}\Lambda^{({-})}\Lambda^{({+})}{=}0$ and $[\Lambda^{({\pm})}]^2{=}\Lambda^{({\pm})}$.

Let us consider the ladder diagram of the $n$-th order in $\gamma_\alpha^{(\textsf{a})}$ shown in Fig. \ref{Fig:Ladder}. The corresponding contribution to the two-particle correlation function \eqref{eq:2PC} is given by the following expression,
 \begin{gather}
\sum_{\sigma_j{=}\pm} \sum_{\mu_j \nu_j} \sum_{\textsf{a}_j,\textsf{b}_j} \int\limits_{\bm{k_j}} 
Y_{\textsf{a}_1\textsf{b}_1}^{(\sigma_1)}(\bm{k_1}) \dots Y_{\textsf{a}_{n-1}\textsf{b}_{n-1}}^{(\sigma_{n-1})}(\bm{k_{n-1}}) \prod_{j=1}^n \sum_{\textsf{c}_j,\alpha_j}\notag \\
{\times} 
  \frac{\gamma_{\alpha_j}^{(\textsf{c}_j)}}{2} \Biggl \{ P_{\mu_j\nu_j} [\overline{\mathcal{L}}^{(\textsf{c}_j,\alpha_j)}_{\bm{k_{(j-1)+}}\bm{k_{j+}}}]_{\textsf{a}_{j-1}\textsf{a}_j} [\mathcal{L}^{(\textsf{c}_j\alpha_j)}_{\bm{k_{j-}}\bm{k_{(j-1)-}}}]_{\textsf{b}_{j} \textsf{b}_{j-1}}  \notag \\
 {+}P_{\nu_j\mu_j}[\mathcal{L}^{(\textsf{c}_j\alpha_j)}_{\bm{k_{(j-1)+}}\bm{k_{j+}}}]_{\textsf{a}_{j-1}\textsf{a}_j} [\overline{\mathcal{L}}^{(\textsf{c}_j\alpha_j)}_{\bm{k_{j-}}\bm{k_{(j-1)-}}}]_{\textsf{b}_j \textsf{b}_{j-1}} \Biggr \}
\notag\\
{\times}
\Bigl [ {\mathcal{G}}_{\bm{p}_+,\textsf{a}}(\varepsilon_+) 
\tau_{\mu_1} \prod_{j=1}^{n-1}  \left (
 \Lambda_{\textsf{a}_j}^{(\sigma_j)} 
  \tau_{\mu_{j+1}}\right ) {\mathcal{G}}_{\bm{q}_+,\textsf{a}^\prime}(\omega_+)\Bigr ]^{\nu\nu^\prime}
\notag \\
{\times} 
\Bigl [ {\mathcal{G}}^T_{\bm{p}_-,\textbf{b}}(\varepsilon_-) \tau_{\nu_1} 
\prod_{j=1}^{n-1}  \left ( [\Lambda_{\textbf{b}_j}^{(-\sigma_j)}]^T 
\tau_{\nu_{j+1}} \right ) {\mathcal{G}}^T_{\bm{q}_-,\textbf{b}^\prime}(\omega_-)\Bigr ]^{\mu\mu^\prime} .
 \label{eq:res:n}
 \end{gather}
 Here we introduced $\omega_\pm {=}\omega{\pm} \Omega/2$ and
$\textsf{a}_{0}{=}\textsf{a}$, $\textsf{a}_n{=}\textsf{a}^\prime$, $\textsf{b}_0{=}\textsf{b}$, $\textsf{b}_n{=}\textsf{b}^\prime$, $\bm{k_0}{=}\bm{p}$, $\bm{k_n}{=}\bm{q}$. Also, we define 
 \begin{equation}
 \begin{split}
 Y^{(+)}_{\textsf{a}\textsf{b}}(\bm{k}) & = \int_E \mathcal{G}_{\bm{k_+},\textsf{a}}^R(E_+)\mathcal{G}_{\bm{k_-},\textsf{b}}^A(E_-), \\
 Y^{(-)}_{\textsf{a}\textsf{b}}(\bm{k}) &= \int_E \mathcal{G}_{\bm{k_+},\textsf{a}}^A(E_+)\mathcal{G}_{\bm{k_-},\textsf{b}}^R(E_-) .
 \end{split}
\end{equation}
We emphasize that for computation of the diffuson ladder in the considered problem one has to integrate over intermediate energies while in the case of disordered fermions, it is not needed since the energy is conserved during scattering on impurity potential. Also, we note other difference between Eq. \eqref{eq:res:n} and a diffuson ladder in the case of impurity scattering. In the former case, the scattering on boson field has a non-trivial matrix structure in the Keldysh space such that the diffuson ladder is sensitive to the distribution function. In the case of impurity scattering, the Green's function causality is preserved and the Keldysh component of the Green's function is not involved. 

Using the structure of the projectors $\Lambda_\textsf{a}^{(\sigma)}$, the matrix $P$, and the matrices $\mathcal{L}$ and $\overline{\mathcal{L}}$, we find that the expression \eqref{eq:res:n} can be dramatically simplified and occurs to be nonzero for $\textsf{a}{=}\textsf{a}^\prime{=}\textsf{b}{=}\textsf{b}^\prime$ only. In that case, it reads
\begin{align}
 2^n &{\mathcal{G}}^R_{\bm{p}_+,\textsf{a}}(\varepsilon_+)
 {\mathcal{G}}^A_{\bm{p}_-,\textsf{a}}(\varepsilon_-)
 {\mathcal{G}}^R_{\bm{q}_+,\textsf{a}}(\omega_+)
 {\mathcal{G}}^A_{\bm{q}_-,\textsf{a}}(\omega_-) ( \Lambda_\textsf{a}^{({+})})^{\nu\nu^\prime}
 \notag \\
 & {\times}
(\Lambda_\textsf{a}^{({-})})^{\mu^\prime\mu}
  \int_{\bm{k_j}} 
Y_{\textsf{a}\textsf{a}}^{(+)}(\bm{k}_1) \dots Y_{\textsf{a}\textsf{a}}^{(+)}(\bm{k}_{n-1})\notag\\
& {\times} \prod_{j=1}^n 
v_{\bm{k_{(j-1)+}}} v^*_{\bm{k_{(j-1)-}}} [U^\dag_{\bm{k_{j+}}} \hat\gamma^{(\textsf{a})} U_{\bm{k_{j-}}}]_{\textsf{a}\textsf{a}}.
 \end{align}
Performing summation over all ladder diagrams from $n{=}1$ to $n{=}\infty$, we find the following result for the ladder
\begin{gather}
2 v_{\bm{p_+}} v^*_{\bm{p_-}}{\mathcal{G}}^R_{\bm{p_+},\textsf{a}}(\varepsilon_+)
 {\mathcal{G}}^A_{\bm{p_-},\textsf{a}}(\varepsilon_-)
 [U^\dag_{\bm{q_+}} \hat\gamma^{(\textsf{a})} U_{\bm{q_-}}]_{\textsf{a}\textsf{a}}
 {\mathcal{G}}^R_{\bm{q_+},\textsf{a}}(\omega_+)
  \notag \\
 {\times}
{\mathcal{G}}^A_{\bm{q_-},\textsf{a}}(\omega_-)
 ( \Lambda_\textsf{a}^{({+})})^{\nu\nu^\prime}(\Lambda_\textsf{a}^{({-})})^{\mu^\prime\mu}\frac{1}{1{-}f_\textsf{a}(\bm{Q},\Omega)} ,
\end{gather}
where 
\begin{gather}
f_\textsf{a}(\bm{Q},\Omega) = 2 \int_{\bm{k}} Y_{\textsf{a}\textsf{a}}^{(+)}(\bm{k})
v_{\bm{k_+}} v^*_{\bm{k_-}} [U^\dag_{\bm{k_+}} \hat\gamma^{(\textsf{a})} U_{\bm{k_-}}]_{\textsf{a}\textsf{a}}.
 \end{gather}
Evaluating $Y_{\textsf{a}\textsf{a}}^{(+)}(\bm{k})$ with the help of Eq. \eqref{eq:G:SCBA}, we find
\begin{gather}
f_\textsf{a}(\bm{Q},\Omega) {=}  \int\limits_{\bm{k}} \frac{2i v_{\bm{k}_+} v^*_{\bm{k}_-} [U^\dag_{\bm{k}_{+}} \hat\gamma^{(\textsf{a})} U_{\bm{k}_{-}}]_{\textsf{a}\textsf{a}}}{\Omega  {-}s_\textsf{a} \xi_{\bm{k}_+}{+} s_\textsf{a} \xi_{\bm{k}_-}{+}i\bar{\gamma}_\textsf{a}(|v_{\bm{k_+}}|^2{+}|v_{\bm{k_-}}|^2)} .
\label{eq:f:res:main}
\end{gather}
Setting $\bm{Q}{=}\Omega{=}0$, we obtain $f_\textsf{a}(0,0){\equiv} 1$, i.e. existence of the pole in the two-particle correlation function in the ladder approximation. Such a pole implies that the corresponding two-particle excitations spread over long distances. 
Expanding the function $f_\textsf{a}(\bm{Q},\Omega)$ in $\bm{Q}$ and $\Omega$, we find
 \begin{gather}
 \frac{1}{1{-}f_\textsf{a}(\bm{Q},\Omega)} \simeq 
 \frac{2 \bar{\gamma}_\textsf{a}^2 /\int_{\bm{k}}([U^\dag_{\bm{k}} \hat\gamma^{(\textsf{a})} U_{\bm{k}}]_{\textsf{a}\textsf{a}}/|v_{\bm{k}}|^2)}
 {D^{(\textsf{a})}_{jl}Q_jQ_l-i\Omega} .
 \label{eq:Diff:f}
 \end{gather}
The matrix of diffusion coefficients $D_{jk}^{(\textsf{a})}$ is given by a lengthy expression in general case. {\color{black} We note that following Eq. \eqref{eq:f:res:main} the diffusion coefficient vanishes for trivial models with all $\xi_{\bm{k}}$, $U_{\bm{k}}$, and $v_{\bm{k}}$ being independent of the momentum $k$. If one of these quantities depend on $k$ the diffusion coefficient is non zero.}

In this paper, we present the expression for $D_{jk}^{(\textsf{a})}$ under the following simplified assumptions: (i) the function $v_{\bm{k}}$ is real and depends on $|\bm{k}|$ only; (ii) the matrix $\hat\gamma^{(\textsf{a})}{=}(\bar{\gamma}_a/n){\rm diag}\{1,1\}$, where $n{=}\int_{\bm{k}}$ is the total particle density; (iii) the non-Abelian vector potential in the momentum space (Berry connection), $\mathcal{A}_j{=}i U^\dag_{\bm{k}} \partial_{\bm{k_j}} U_{\bm{k}}$, satisfies the condition $\int_{\bm{k}} \mathcal{A}_j{=}0$. Under such assumptions, the diffusion coefficients become 
\begin{gather}
D^{(\textsf{a})}_{jl} = \frac{1}{2\int\limits_{\bm{k}} |v_k|^{-2}}  \int_{\bm{k}} \Biggl \{
 \frac{\delta_{jl}}{\bar{\gamma}_a d}
\left [ \frac{(\nabla_{\bm{k}} \xi_k)^2}{|v_k|^4} + 2 \bar{\gamma}_a^2\frac{(\nabla_{\bm{k}} v_k)^2}{v_k^2}\right ] \notag \\
{+} \Bigl [  \bar{\gamma}_\textsf{a}(\mathcal{A}_j \mathcal{A}_l{+}\mathcal{A}_l \mathcal{A}_j)
{-} \frac{s_a}{v_k^2} (\mathcal{A}_j \partial_{\bm{k_l}}  \xi_k{+}\mathcal{A}_l\partial_{\bm{k_j}}  \xi_k) \Bigr ]_{\textsf{a}\textsf{a}} \Bigr \} .
\label{eq:res:D}
\end{gather}
It is worthwhile to mention that a nonzero diffusion coefficient, ${\propto}\bar{\gamma}_\textsf{a}$,  appears even in the cases of either a flat band $\xi_q{=}{\rm const}$ or in the absence of Hamiltonian,  $\xi_q{=}0$. {\color{black} We note that the first term in the r.h.s. of Eq. \eqref{eq:res:D} corresponds to a standard scenario in which diffusion is determined by the spectrum curvature. The second term in the r.h.s. of Eq. \eqref{eq:res:D} describes the contribution to the diffusion coefficient from dispersion of parameter $v_{\bm{k}}$ controlling non-locality of scattering. The third contribution to $D_{jl}^{\textsf{a}}$ involves the non-Abelian vector potential in the combination resembling the quantum metric tensor.}

\noindent\textsf{\color{blue} Example. --- } To illustrate the general result \eqref{eq:res:D} we apply it to the model of two-band Chern insulator with the Chern number equal $-1$ proposed in Ref.~\cite{Tonielli2020}. The Hamiltonian of that model is 
$H_0(\bm{q}){=}\bm{d}_{\bm{q}} {\cdot} \bm{\sigma}$, where $\bm{d}_{\bm{q}}{=}\{2m q_x,2m q_y,q^2{-}m^2\}$. Consequently,  we find $\xi_q{=}d_q{=}q^2{+}m^2$  and $U_{\bm{q}}{=}(q_x {-} iq_y\sigma_z{-}im \sigma_y)/\sqrt{d_q}$. Also, we choose $v_q{=}\sqrt{d_q}$ that makes the relation between fermionic fields $l_\textsf{a}$ and $\psi_\alpha$ to be local in space. Then, using Eq. \eqref{eq:res:D}, we obtain $D_{jl}^{(\textsf{a})}{=}D_{\textsf{a}}\delta_{jl}$, where 
\begin{gather}
D_{\textsf{a}}= \frac{2}{d \bar{\gamma}_\textsf{a}} \frac{\int_{\bm{k}} k^2/d_k^2}{\int_{\bm{k}}1/d_k} +\bar{\gamma}_\textsf{a} .
\end{gather}
For $\bar{\gamma}_\textsf{u}{=}\bar{\gamma}_\textsf{d}$ the above expression has been originally derived in Ref. \cite{Nosov2023}.

\noindent\textsf{\color{blue} Discussion. --- } The result of self-consistent Born approximation for the single-particle Green's function suggests that the relevant time scale for excitations in our system is of the order of  $1/(\bar{\gamma}_{\textsf{a}} m^2)$. However, similarly to the disordered systems, there  is typically a much longer time which determines spreading of the particle density. Indeed, the two-particle correlation function \eqref{eq:2PC} can be considered as Green's function for the linear equation governing time and spatial dynamics of the deviation of the particle density $\delta n_{\textsf{a}}(\bm{x},t)$ from the dark state with $n_{\textsf{d}}{=}1{-}n_{\textsf{u}}{=}n$ (see Ref. \cite{Nosov2023}). Our result implies that $\delta n_{\textsf{a}}(\bm{x},t)$ obeys the diffusion equation. Since the diffusion equations for $\delta n_{\textsf{u}}(\bm{x},t)$ and $\delta n_{\textsf{d}}(\bm{x},t)$ are independent, the diffusion can spatially redistribute the $\textsf{u}$- and $\textsf{d}$-particles within a given band only. In particular, if one creates a perturbation of particle densities, they will spread over a system of size $L$ for time ${\sim} L^2/D_{\textsf{a}}$. However, there exists a recombination between $\textsf{u}$-particles and $\textsf{d}$-holes that results in a nonlinear term, ${\propto}  \delta n_{\textsf{u}} \delta n_{\textsf{d}}$, that couples  the diffusion equations. An accurate derivation of the recombination contribution for the considered general model is beyond the scope of our work. We just mention that  the recombination results in a power law decay of density perturbation from the dark state (see Ref. \cite{Nosov2023} for details). Such slow decay can obviously complicate engineering of the desired dark state in  a real setup. 

Other effect which is also beyond the scope of our paper is instability of the dark state due to pumping of particles into $\textsf{u}$-band, predicted in Ref. \cite{Nosov2023} for the model of Ref. \cite{Tonielli2020}. In our approach, such instability of the dark state should appear after inclusion of self-energy diagrams to the diffuson ladder as a modification of the denominator of \eqref{eq:Diff:f}, $D_{jl}^{(\textsf{a})}Q_jQ_l{-}i\Omega {\to} D_{jl}^{(\textsf{a})}Q_jQ_l{-}i\Omega{+}1/\tau_{\phi}^{(\textsf{a})}$. 
The nonzero dephasing rate of diffuson is possible since the diffusion pole is not preserved by conservation of $\textsf{u}$- or $\textsf{d}$-fermions separately. The negative sign of the dephasing rate, $1/\tau_{\phi}{<}0$, would break causality and indicate instability of the dark state. 

In addition to the appearance of the dephasing rate, there could be corrections (of weak-localization-type) to the diffusion coefficient found within a ladder approximation. A source of such corrections is the momentum dependence of diffuson self-energy. Some of the corresponding diagrams can be recast in the form of interaction of several diffusons. As known from treatment of the disordered systems, such diagrams can be conveniently summed by means of the nonlinear sigma model. For the model of spinless (single-band) fermions subjected to random measurements, such nonlinear sigma models have been recently derived in Refs. \cite{Yang2022,Fava2023,Poboiko2023}. One can also study the distribution function of $Z[\Phi,\overline{\Phi}]$ with the help of the nonlinear sigma model \cite{Poboiko2023}. It is a challenge to derive a nonlinear sigma model for the generalized model considered in this paper. {\color{black} Also it could be interesting to extend our model by adding elastic scattering in $H_0$. Then similar to Ref. \cite{Pastawski} one can study the interplay of elastic and dissipative scattering in the diffusion coefficient.}

{\color{black} Finally, we mention that our results for diffusion behavior is different from the ones in Refs. \cite{Esposito2005,Castro-Alvaredo2016,Dhar2019,Jin2022} in the following ways: (i) our consideration is not restricted to 1D models; (ii) we demonstrate that diffusion emerges even in the absence of spectrum dispersion; (iii) we elucidate the physical origin of the diffusion as correlated propagation of electron-hole pairs in each band.}

\noindent\textsf{\color{blue} Summary. --- } To summarize, we studied the emergence of the diffusive excitations in the generalized two-band dissipative 
quantum many-body state preparation dynamics, which conserves the total number of particles. We derive the general expression for the diffusion coefficient that determines the diffusion pole in the diffuson ladder for intra-band particle-hole excitations.
In the presence of the band dispersion and at $|\bar{\gamma}_\textsf{a}|{\ll}1$, the diffusion coefficient is inversely proportional to the scattering (dissipation) rate as expected. In the case of a flat band or in the absence of the Hamiltonian part, the diffusion coefficient is still nonzero and proportional to the dissipation rate. {\color{black} Therefore, our analysis shows that intra-band diffusion emerges generically in the number-conserving dissipative systems described by GSKL master equation.} In contrast, the inter-band two-particle excitations are not diffusive. They decay on the time scale determined by the single-particle decay rate $\bar{\gamma}_\textsf{a} \xi_q$. Our work opens up many future research directions. 

\noindent\textsf{\color{blue} Acknowledgments. --- } We thank A. Altland, S. Diehl, M. Glazov, I. Gornyi, I. Poboiko, and D. Smirnov for useful discussions. I.S.B. is grateful to M. Goldstein, P. Nosov, and D. Shapiro for collaboration on a related project. The work was funded by the Russian Science Foundation under the Grant No. 22-22-00641.

\end{document}